\documentclass{article}
\usepackage{url}
\usepackage{color}
\usepackage{bm}
\usepackage{amsmath}
\usepackage{amssymb}\usepackage{amsthm}
\usepackage{amsfonts}
\usepackage{amscd}
\begin{document}
\theoremstyle{plain}
\newtheorem*{ithm}{Theorem}
\newtheorem*{idefn}{Definition}
\newtheorem{thm}{Theorem}[section]
\newtheorem{lem}[thm]{Lemma}
\newtheorem{dlem}[thm]{Lemma/Definition}
\newtheorem{prop}[thm]{Proposition}
\newtheorem{set}[thm]{Setting}
\newtheorem{cor}[thm]{Corollary}
\newtheorem*{icor}{Corollary}
\theoremstyle{definition}
\newtheorem{assum}[thm]{Assumption}
\newtheorem{notation}[thm]{Notation}
\newtheorem{defn}[thm]{Definition}
\newtheorem{clm}[thm]{Claim}
\newtheorem{ex}[thm]{Example}
\theoremstyle{remark}
\newtheorem{rem}[thm]{Remark}
\numberwithin{equation}{section}

\newcommand{\unit}{\mathbb I}
\newcommand{\ali}[1]{{\mathfrak A}_{[ #1 ,\infty)}}
\newcommand{\alm}[1]{{\mathfrak A}_{(-\infty, #1 ]}}
\newcommand{\nn}[1]{\lV #1 \rV}
\newcommand{\br}{{\mathbb R}}
\newcommand{\dm}{{\rm dom}\mu}
\newcommand{\inn}{({\rm {inner}})}
\newcommand{\Ad}{\mathop{\mathrm{Ad}}\nolimits}
\newcommand{\Proj}{\mathop{\mathrm{Proj}}\nolimits}
\newcommand{\RRe}{\mathop{\mathrm{Re}}\nolimits}
\newcommand{\RIm}{\mathop{\mathrm{Im}}\nolimits}
\newcommand{\Wo}{\mathop{\mathrm{Wo}}\nolimits}
\newcommand{\Prim}{\mathop{\mathrm{Prim}_1}\nolimits}
\newcommand{\Primz}{\mathop{\mathrm{Prim}}\nolimits}
\newcommand{\ClassA}{\mathop{\mathrm{ClassA}}\nolimits}
\newcommand{\Class}{\mathop{\mathrm{Class}}\nolimits}
\newcommand{\diam}{\mathop{\mathrm{diam}}\nolimits}
\def\qed{{\unskip\nobreak\hfil\penalty50
\hskip2em\hbox{}\nobreak\hfil$\square$
\parfillskip=0pt \finalhyphendemerits=0\par}\medskip}
\def\proof{\trivlist \item[\hskip \labelsep{\bf Proof.\ }]}
\def\endproof{\null\hfill\qed\endtrivlist\noindent}
\def\proofof[#1]{\trivlist \item[\hskip \labelsep{\bf Proof of #1.\ }]}
\def\endproofof{\null\hfill\qed\endtrivlist\noindent}

\newcommand{\qe}{\sim_{\mathop{q.e.}}}
\newcommand{\bgu}[2]{\beta_{#1}^{U#2}}
\newcommand{\SDC}{{\mathfrak A}_{\mathop{\mathrm SDC}}}
\newcommand{\sdc}{\SDC}
\newcommand{\mkA}{{\mathfrak A}}
\newcommand{\mkB}{{\mathfrak B}}
\newcommand{\mkC}{{\mathfrak C}}
\newcommand{\mkD}{{\mathfrak D}}
\newcommand{\mkh}{{\mathfrak h}}
\newcommand{\mkk}{{\mathfrak K}}
\newcommand{\mkK}{{\mathfrak K}}
\newcommand{\varphii}{\varphi}
\newcommand{\sdcizs}{\sdc\lmk \mkk_1\oplus \mkk_0\oplus \mkk_3,  \mkC_1\oplus \mkC_0\oplus \mkC_3\rmk}
\newcommand{\sdciz}{\sdc\lmk \mkk_1\oplus \mkk_0,  \mkC_1\oplus \mkC_0\rmk}
\newcommand{\sdcin}{\sdc\lmk \mkk_1\oplus \mkk_2,  \mkC_1\oplus \mkC_2\rmk}
\newcommand{\sdcis}{\sdc\lmk \mkk_1\oplus \mkk_3,  \mkC_1\oplus \mkC_3\rmk}
\newcommand{\sdckcz}[1]{\sdc^{(0)}\lmk\mkk_{#1},\mkC_{#1}\rmk}
\newcommand{\sdckc}[1]{\sdc\lmk\mkk_{#1},\mkC_{#1}\rmk}
\newcommand{\bh}[1]{\caB\lmk\caH_{#1}\rmk}
\newcommand{\pg}{{\mathfrak S}(\bbZ^2)}
\newcommand{\oo}{{\boldsymbol\varphii}}
\newcommand{\caA}{{\mathcal A}}
\newcommand{\caB}{{\mathcal B}}
\newcommand{\caC}{{\mathcal C}}
\newcommand{\caD}{{\mathcal D}}
\newcommand{\caE}{{\mathcal E}}
\newcommand{\caF}{{\mathcal F}}
\newcommand{\caG}{{\mathcal G}}
\newcommand{\caH}{{\mathcal H}}
\newcommand{\caI}{{\mathcal I}}
\newcommand{\caJ}{{\mathcal J}}
\newcommand{\caK}{{\mathcal K}}
\newcommand{\caL}{{\mathcal L}}
\newcommand{\caM}{{\mathcal M}}
\newcommand{\caN}{{\mathcal N}}
\newcommand{\caO}{{\mathcal O}}
\newcommand{\caP}{{\mathcal P}}
\newcommand{\caQ}{{\mathcal Q}}
\newcommand{\caR}{{\mathcal R}}
\newcommand{\caS}{{\mathcal S}}
\newcommand{\caT}{{\mathcal T}}
\newcommand{\caU}{{\mathcal U}}
\newcommand{\caV}{{\mathcal V}}
\newcommand{\caW}{{\mathcal W}}
\newcommand{\caX}{{\mathcal X}}
\newcommand{\caY}{{\mathcal Y}}
\newcommand{\caZ}{{\mathcal Z}}
\newcommand{\bbA}{{\mathbb A}}
\newcommand{\bbB}{{\mathbb B}}
\newcommand{\bbC}{{\mathbb C}}
\newcommand{\bbD}{{\mathbb D}}
\newcommand{\bbE}{{\mathbb E}}
\newcommand{\bbF}{{\mathbb F}}
\newcommand{\bbG}{{\mathbb G}}
\newcommand{\bbH}{{\mathbb H}}
\newcommand{\bbI}{{\mathbb I}}
\newcommand{\bbJ}{{\mathbb J}}
\newcommand{\bbK}{{\mathbb K}}
\newcommand{\bbL}{{\mathbb L}}
\newcommand{\bbM}{{\mathbb M}}
\newcommand{\bbN}{{\mathbb N}}
\newcommand{\bbO}{{\mathbb O}}
\newcommand{\bbP}{{\mathbb P}}
\newcommand{\bbQ}{{\mathbb Q}}
\newcommand{\bbR}{{\mathbb R}}
\newcommand{\bbS}{{\mathbb S}}
\newcommand{\bbT}{{\mathbb T}}
\newcommand{\bbU}{{\mathbb U}}
\newcommand{\bbV}{{\mathbb V}}
\newcommand{\bbW}{{\mathbb W}}
\newcommand{\bbX}{{\mathbb X}}
\newcommand{\bbY}{{\mathbb Y}}
\newcommand{\bbZ}{{\mathbb Z}}
\newcommand{\str}{^*}
\newcommand{\lv}{\left \vert}
\newcommand{\rv}{\right \vert}
\newcommand{\lV}{\left \Vert}
\newcommand{\rV}{\right \Vert}
\newcommand{\la}{\left \langle}
\newcommand{\ra}{\right \rangle}
\newcommand{\ltm}{\left \{}
\newcommand{\rtm}{\right \}}
\newcommand{\lcm}{\left [}
\newcommand{\rcm}{\right ]}
\newcommand{\ket}[1]{\lv #1 \ra}
\newcommand{\bra}[1]{\la #1 \rv}
\newcommand{\lmk}{\left (}
\newcommand{\rmk}{\right )}
\newcommand{\al}{{\mathcal A}}
\newcommand{\md}{M_d({\mathbb C})}
\newcommand{\eaut}{\mathop{\mathrm{EAut}}\nolimits}
\newcommand{\qaut}{\mathop{\mathrm{QAut}}\nolimits}
\newcommand{\sqaut}{\mathop{\mathrm{SQAut}}\nolimits}
\newcommand{\gsqaut}{\mathop{\mathrm{GSQAut}}\nolimits}
\newcommand{\QLS}{\mathop{\mathcal{SL}}\nolimits}
\newcommand{\haut}{\mathop{\mathrm{HAut}}\nolimits}
\newcommand{\guaut}{\mathop{\mathrm{GUQAut}}\nolimits}
\newcommand{\IG}{\mathop{\mathrm{IG}}\nolimits}
\newcommand{\IP}{\mathop{\mathrm{IP}}\nolimits}
\newcommand{\ainn}{\mathop{\mathrm{AInn}}\nolimits}
\newcommand{\id}{\mathop{\mathrm{id}}\nolimits}
\newcommand{\Tr}{\mathop{\mathrm{Tr}}\nolimits}
\newcommand{\co}{\mathop{\mathrm{co}}\nolimits}
\newcommand{\sym}{\mathop{\mathrm{Sym}}\nolimits}
\newcommand{\sgn}{\mathop{\mathrm{sgn}}\nolimits}
\newcommand{\Ran}{\mathop{\mathrm{Ran}}\nolimits}
\newcommand{\Ker}{\mathop{\mathrm{Ker}}\nolimits}
\newcommand{\Aut}{\mathop{\mathrm{Aut}}\nolimits}
\newcommand{\QAut}{\mathop{\mathrm{QAut}}\nolimits}
\newcommand{\EAut}{\mathop{\mathrm{EAut}}\nolimits}
\newcommand{\SPT}{\mathop{\mathrm{SPT}}\nolimits}
\newcommand{\Inn}{\mathop{\mathrm{Inn}}\nolimits}
\newcommand{\spn}{\mathop{\mathrm{span}}\nolimits}
\newcommand{\Mat}{\mathop{\mathrm{M}}\nolimits}
\newcommand{\Dia}{\mathop{\mathrm{D}}\nolimits}
\newcommand{\UT}{\mathop{\mathrm{UT}}\nolimits}
\newcommand{\DT}{\mathop{\mathrm{DT}}\nolimits}
\newcommand{\GL}{\mathop{\mathrm{GL}}\nolimits}
\newcommand{\spa}{\mathop{\mathrm{span}}\nolimits}
\newcommand{\supp}{\mathop{\mathrm{supp}}\nolimits}
\newcommand{\rank}{\mathop{\mathrm{rank}}\nolimits}
\newcommand{\idd}{\mathop{\mathrm{id}}\nolimits}
\newcommand{\ran}{\mathop{\mathrm{Ran}}\nolimits}
\newcommand{\dr}{ \mathop{\mathrm{d}_{{\mathbb R}^k}}\nolimits} 
\newcommand{\dc}{ \mathop{\mathrm{d}_{\cc}}\nolimits} \newcommand{\drr}{ \mathop{\mathrm{d}_{\rr}}\nolimits} 
\newcommand{\auz}{\Aut^{(0)}}
\newcommand{\zin}{\mathbb{Z}}
\newcommand{\rr}{\mathbb{R}}
\newcommand{\cc}{\mathbb{C}}
\newcommand{\ww}{\mathbb{W}}
\newcommand{\nan}{\mathbb{N}}\newcommand{\bb}{\mathbb{B}}
\newcommand{\aaa}{\mathbb{A}}\newcommand{\ee}{\mathbb{E}}
\newcommand{\pp}{\mathbb{P}}
\newcommand{\wks}{\mathop{\mathrm{wk^*-}}\nolimits}
\newcommand{\Hom}{\mathop{\mathrm{Hom}}\nolimits}
\newcommand{\mk}{{\Mat_k}}
\newcommand{\mnz}{\Mat_{n_0}}
\newcommand{\mn}{\Mat_{n}}
\newcommand{\dist}{\mathrm{d}}
\newcommand{\braket}[2]{\left\langle#1,#2\right\rangle}
\newcommand{\ketbra}[2]{\left\vert #1\right \rangle \left\langle #2\right\vert}
\newcommand{\abs}[1]{\left\vert#1\right\vert}
\newtheorem{nota}{Notation}[section]
\def\qed{{\unskip\nobreak\hfil\penalty50
\hskip2em\hbox{}\nobreak\hfil$\square$
\parfillskip=0pt \finalhyphendemerits=0\par}\medskip}
\def\proof{\trivlist \item[\hskip \labelsep{\bf Proof.\ }]}
\def\endproof{\null\hfill\qed\endtrivlist\noindent}
\def\proofof[#1]{\trivlist \item[\hskip \labelsep{\bf Proof of #1.\ }]}
\def\endproofof{\null\hfill\qed\endtrivlist\noindent}
\newcommand{\pa}[1]{\pi_{#1}\alpha_{#1}}
\newcommand{\pat}{\lmk \pa L\hat\otimes \pa R\rmk}
\newcommand{\ao}[1]{a_\omega(#1)}
\newcommand{\eg}[2]{(-1)^{\ao{#1}}#2}
\newcommand{\unic}{\unit_{\bbC^2}}
\newcommand{\unih}[1]{\unit_{\caH_{#1}}}
\newcommand{\ZZ}{\bbZ_2\times\bbZ_2}
\newcommand{\SSS}{\mathcal{S}}
\newcommand{\cs}{S}
\newcommand{\ct}{t}
\newcommand{\hS}{S}
\newcommand{\vv}{{\boldsymbol v}}
\newcommand{\ala}{a}
\newcommand{\bet}{b}
\newcommand{\gam}{c}
\newcommand{\alphas}{\alpha}
\newcommand{\alphai}{\alpha^{(\sigma_{1})}}
\newcommand{\alphan}{\alpha^{(\sigma_{2})}}
\newcommand{\betas}{\beta}
\newcommand{\betai}{\beta^{(\sigma_{1})}}
\newcommand{\betan}{\beta^{(\sigma_{2})}}
\newcommand{\alphass}{\alpha^{{(\sigma)}}}
\newcommand{\uu}{V}
\newcommand{\vp}{\varsigma}
\newcommand{\vpr}{R}
\newcommand{\tg}{\tau_{\Gamma}}
\newcommand{\sgg}{\Sigma_{\Lambda}}
\newcommand{\nh}{1}
\newcommand{\rk}{2,a}
\newcommand{\nii}{1,a}
\newcommand{\nhh}{3,a}
\newcommand{\sjt}{2}
\newcommand{\sjtg}{2}
\newcommand{\bcg}{\caB(\caH_{\alpha})\otimes  C^{*}(\Sigma_{\Lambda})}
\newcommand{\cacn}[2]{\caA_{C_{#1}\cap H_{#2}^{{c_{#2}^{(2)}}}}}
\newcommand{\caci}[2]{\caA_{C_{#1}\cap H_{#2}^{{c_{#2}^{(1)}}}}}
\newcommand{\cac}[2]{\caA_{C_{#1}\cap H_{#2}^{c_{#2}}}}
\newcommand{\cacc}[3]{\caA_{\lmk C_{#1}\cap H_{#2}^{c^{(#3)}_{#2}} \rmk^c\cap H_{#2} }}
\newcommand{\Uo}{{\mathrm U}(1)}
\newcommand{\css}[2]{c_{#1}^{#2}}
\newcommand{\uss}[2]{u_{#1}^{#2}}
\newcommand{\kss}[2]{\kappa_{#1}^{#2}}
\newcommand{\mss}[3]{m_{#1#2}^{#3}}

\newcommand{\PD}{\mathcal{PD}(G)}
\newcommand{\PDz}{\mathcal{PD}_0(G)}
\newcommand{\ptp}{\lmk\pi_L\hat\otimes\pi_R\rmk}
\newcommand{\bebe}[3]{\bgu{#2}#1\lmk\eta_{#2#1}^\epsilon\rmk^{-1}
\bgu{#3}#1\lmk \lmk \eta_{#3#1}^{\eg{#2}\epsilon}\rmk_{-\ao{#2}\epsilon}\rmk^{-1}
\eta_{#2#3,#1}^\epsilon \lmk \bgu{#2#3}#1\rmk^{-1}}
\newcommand{\etas}[3]{\eta_{#1#2}^{#3}}
\newcommand{\os}[1]{\omega_{#1}}
\newcommand{\bebey}[4]{\bgu{#2}#1\lmk\eta_{#2#1}^{\epsilon #4}\rmk^{-1}
\bgu{#3}#1\lmk \lmk \eta_{#3#1}^{\eg{#2}{\epsilon #4}}\rmk_{-\ao{#2}{\epsilon }}\rmk^{-1}
\eta_{#2#3,#1}^{\epsilon #4} \lmk \bgu{#2#3}#1\rmk^{-1}}
\newcommand{\tdnh}{\delta_{1}}
\newcommand{\dnh}{\delta_{\nhh}}
\newcommand{\eijz}[2]{e_{#1#2}^{(\Lambda_0)}}
\newcommand{\tdsj}{\delta_{2}}
\newcommand{\tgsj}{\tilde\caG_{1}}
\newcommand{\dn}{\delta_{\nii}}
\newcommand{\tdnhs}{\delta_{3}}
\newcommand{\tgsjz}{\tilde \caG_2}
\newcommand{\tdsjs}{\delta_4}
\newcommand{\sigman}{i}
\newcommand{\Rn}{2}
\newcommand{\Ln}{1}
\newcommand{\dro}{\delta_{\rk}}
\newcommand{\che}[1]{#1}

\title{$2$-d Fermionic SPT with CRT symmetry}

\author{Yoshiko Ogata \thanks{ Graduate School of Mathematical Sciences
The University of Tokyo, Komaba, Tokyo, 153-8914, Japan
Supported in part by
the Grants-in-Aid for
Scientific Research, JSPS.}}
\maketitle
\begin{abstract}
An invariant of
SPT-phases with on-site finite group $G$ symmetry for two-dimensional Fermion systems
is derived in \cite{2df}.
This invariant is doubled compared to the conjectured one from the invertible quantum field theory.
We show that if we require CRT-symmetry (which holds automatically in quantum field theory) 
in addition, then our invariant reduces to the conjectured one.
\end{abstract}

\section{Introduction}\label{introsec}
The notion of symmetry protected topological (SPT) phases was introduced by Gu and Wen [GW].
It is defined as follows:
we consider the set of all Hamiltonians with some symmetry, 
which have a unique gapped ground state in the bulk, and can be smoothly deformed into
a common trivial gapped Hamiltonian without closing the gap.
We say 
two such Hamiltonians are equivalent, if they can be smoothly deformed into
each other, without closing the gap and without breaking the symmetry.
We call an equivalence class of this classification, a
symmetry protected topological (SPT) phase.
In \cite{2df} we derived some 
invariant of two-dimensional Fermionic SPT-phases
for finite on-site group $G$.

Our invariant of \cite{2df} is
given by
\begin{align}
\lmk C^3(G,\Uo\oplus\Uo)\rmk
\times \lmk H^2(G, \bbZ_2\oplus\bbZ_2)\rmk
\times \lmk H^1(G, \bbZ_2)\rmk,
\end{align}
devided by some equivalence relation.
More precisely, given $a\in H^1(G, \bbZ_2)$,
we define a group action 
on $C^3(G,\Uo\oplus\Uo)$, $ C^2(G, \bbZ_2\oplus\bbZ_2)$.
Let $a\in H^1(G, \bbZ_2)$ and $A:=\bbZ_2$ or $A:=\Uo$.
We define a $G$-action on $A\oplus A$ by 
\begin{align}
G\times \lmk A\oplus A\rmk\ni (g,x)
\mapsto 
 x^{a(g)}
:=\begin{pmatrix}
0&1\\1&0
\end{pmatrix}^{a(g)} x\in A\oplus A.
\end{align}
We associate $A\oplus A$ the point-wise multiplication , i.e.,
for $x=(x_+,x_-), y=(y_+,y_-)\in A\oplus A$, we set $x\cdot y:=(x_+y_+, x_-y_-)$.
For $x=(x_+,x_-)\in \bbZ_2\oplus \bbZ_2$, we also
set $(-1)^x:=((-1)^{x_+}, (-1)^{x_-})\in \Uo\oplus \Uo$.
For $x\in C^1(G,A\oplus A)$, $y\in C^2(G,A\oplus A)$, $z\in C^3(G,A\oplus A)$ and $a\in H^1(G,\bbZ_2)$,
we set
\begin{align}
\begin{split}
&d_a^1 x(g,h):=\frac{\lmk x^{a(g)}(h)\rmk \cdot x(g)}{x(gh)},\\
&d_a^2 y(g,h,k):=\frac{\lmk y^{a(g)}(h,k)\rmk\cdot y(g,hk)} {y(gh,k)\cdot y(g,h)},\\
&d_a^3 z(g,h,k,f):=\frac{\lmk  \lmk z^{a(g)}(h,k,f)\rmk\rmk \cdot z(g, hk, f)\cdot z(g,h,k)}{z(gh,k, f)\cdot z(g,h,kf)}.
\end{split}
\end{align}

We denote by $\widetilde \PDz$ the triple $(c, \kappa,a)$
of 
\begin{align}
c\in C^3\lmk G, \Uo\oplus \Uo\rmk, \;
\kappa\in C^2\lmk G, \bbZ_2\oplus \bbZ_2\rmk,\;
a\in H^1(G,\bbZ_2)
\end{align}
such that
\begin{align}
d_a^2 \kappa(g,h,k)
=0,\quad 
d_a^3 c(g,h,k,f)
=(-1)^{\kappa(g,h)\cdot \lmk \kappa^{a(gh)}(k,f)\rmk}.\label{c49}
\end{align}
On $\widetilde \PDz$
set 
\[
(c^{(1)}, \kappa^{(1)},a^{(1)})\sim_{\PDz}(c^{(2)}, \kappa^{(2)},a^{(2)})
\]
 for $(c^{(1)}, \kappa^{(1)},a^{(1)}), (c^{(2)}, \kappa^{(2)},a^{(2)})\in \widetilde\PDz$
if
the following hold.
\begin{description}
\item[(i)]$a^{(1)}(g)=a^{(2)}(g)=: a(g)$ for any $g\in G$, and
\item[(ii)] there exist an $m\in C^1(G, \bbZ_2\oplus \bbZ_2)$
and a $\sigma\in C^2(G, \Uo\oplus \Uo)$
such that
\begin{align}
&\kss {} {(2)}(g,h)=d_a^1 m(g,h)+\kss {} {(1)}(g,h),\\
&c^{(2)}(g,h,k)
=\lmk-1\rmk^{\kss {} {(1)}(g,h) \cdot  m^{a(gh)}(k)}\lmk -1\rmk^{m(g)
\cdot \lmk \kss {} {(2)}\rmk^{a(g)} (h,k)} d_a^2\sigma(g,h,k) c^{(1)}(g,h,k).
\end{align}
\end{description}
We denote the equivalence class by $\PDz$.
In \cite{2df}, the following Theorem is proven.
\begin{thm}[Theorem 1.5 of \cite{2df}]
There is a $\PDz$-valued index  invariant
$\caT_{\Phi}$
 of the classification
of 2d Fermionic SPT $\Phi\in \caP_{SL\beta}$ (Definition 1.4 \cite{2df}).
\end{thm}
For the concrete microscopic setting and the derivation of
the invariant, see the original paper \cite{2df}.
We use the setting, results, notation from \cite{2df} freely.

Without the doubled structure like $\Uo\oplus \Uo$, our
result in \cite{2df} coincides with the one predicted in the invertible quantum field theory \cite{BM} \cite{WG},
the Pontrjagin dual of $3$-dimensional Spin Bordism on $BG$
$\Hom(\Omega_3^{spin}(BG), \bbR\setminus \bbZ)$.
In particular, the set of the diagonal elements in $\PDz$
\begin{align}\label{pd}
\left\{
[(c, \kappa ,a)]_{\PDz}
\mid 
c^{+1}(g,h,k)=c^{-1}(g,h,k),\;\;
\kappa^{+1}(g,h)=\kappa^{-1}(g,h),\quad g,h,k\in G
\right\}
\end{align}
is isomorphic to $\Hom(\Omega_3^{spin}(BG), \bbR\setminus \bbZ)$.
We denote this set (\ref{pd})
 by $\Hom(\Omega_3^{spin}(BG), \bbR\setminus \bbZ)$.

Recall in quantum field theory, there are symmetries arising automatically from the axiom, that is
CRT symmetry \cite{haag}\cite{FH}.
In particular, there is an anti-linear reflection with respect to the $y$-axis.
 In condensed matter physics, apriori there is no such symmetry.
In this paper, we show if we require such symmetry in our setting,
our index reduces to the predicted one.

More precisely, let $R:\bbZ^2\to\bbZ^2$ be the reflection with respect to the $y$-axis:
\begin{align}
\caR(x,y):=(-x-1,y).
\end{align}
Let $\varsigma$ be  an anti-linear automorphismon  on $\caA$ such that
\begin{align}\label{xip}
\begin{split}
\varsigma^2=\id,\quad \varsigma\beta_g^X=\beta_g^{\caR X}\varsigma,\quad
\varsigma\lmk \caA_X\rmk=\caA_{\caR X},\quad
\varsigma\Theta=\Theta\varsigma,\quad \varsigma\tau=\tau^{-1}\varsigma,\quad
\omega^{(0)}\varsigma(A^*)=\omega^{(0)}(A)
\end{split}
\end{align}
for all $X\subset \bbZ^2$, $g\in G$, $A\in \caA$.
Recall the definition of $\tau$ from (3.17) \cite{2df}
and other automorphisms, states, algebras from section 1 \cite{2df}.
We denote by $\caP_{SL\beta\varsigma}$
the set of all $\Phi\in \caP_{SL\beta}$ (Definition 1.4 \cite{2df})
which are also
$\varsigma$-invariant.
\begin{rem}
For example, let $\hat R : \mathfrak h_{\bbZ^2}\to\mathfrak h_{\bbZ^2}$ be an anti-linear map
such that
\begin{align}
\hat R\delta_{(x,y)}:=\delta_{\caR(x,y)}.
\end{align}
 Let $e_j$, $j=0,\ldots,  d-1$ be the standard basis of $\bbC^d$.
If our $U_g$ satisfies
\begin{align}
\braket{e_j}{U_ge_k}=\braket{e_{d-j-1}}{ U_ge_{d-k-1}}
\end{align}
for all $g\in G$ and $j,k=0,\ldots, d-1$, then
\begin{align}
\varsigma\lmk B(f)\rmk:= B( \hat R f),\quad f\in \mathfrak h_{\bbZ^2}
\end{align}
defines
an anti-linear automorphism on $\caA$ satisfying the condition (\ref{xip}).
There can be many lattice realization of the CRT-symmetry.
Therefore, we do not specify a concrete form of it here.
\end{rem}
The main result of this paper is the following.
\begin{thm}
For $\Phi\in \caP_{SL\beta\varsigma}$,
the $\PDz$-valued index 
$\caT_{\Phi}$
belongs to $\Hom(\Omega_3^{spin}(BG), \bbR\setminus \bbZ)$ (\ref{pd}).
\end{thm}

\section{Proof}
Let $\Phi\in \caP_{SL\beta\varsigma}$ with the unique ground state $\omega$.
From the construction of
the automorphic equivalence,
we can choose
$\alpha\in \EAut(\omega)$ and 
 $(\alpha_L,\alpha_R, \Upsilon)\in \caD^V(\alpha, \theta)$
so that 
\begin{align}
\varsigma\alpha=\alpha\varsigma,\quad\varsigma\alpha_L=\alpha_R\varsigma,\quad
\varsigma\Upsilon=\Upsilon\varsigma.
\end{align}
We can also choose $(\alpha_U,\alpha_D,\Xi_L,\Xi_R)\in \caD^H(\alpha,\theta)$
so that 
\begin{align}
\varsigma\alpha_U=\alpha_U\varsigma,\quad\varsigma\alpha_D=\alpha_D\varsigma,\quad
\varsigma\Xi_L=\Xi_R\varsigma.
\end{align}
Analougous chouce can be made for automorphisms $\alpha_{\partial C_\theta, \sigma,D}$,
$\alpha_{C_\theta, \sigma,D}$ in the proof of Lemma 5.1 \cite{2df}.

Therefore,
in the proof of Lemma 5.1 \cite{2df}, 
we may choose $Y_\sigma$ so that
$
Y_L=\varsigma Y_R \varsigma
$. Therefore, we may assume 
$\varphi_L(A)=\varphi_R\varsigma(A^*)$, for $A\in \caA_{H_L\cap H_U\cap C_{\theta''}}$,
$\varphi_R(A)=\varphi_L\varsigma(A^*)$, for $A\in \caA_{H_R\cap H_U\cap C_{\theta''}}$
for $\varphi_L,\varphi_R$ in the proof of Proposition 5.2 \cite{2df}.
Therefore, $\varphi_L\hat\otimes \varphi_R$ is $\varsigma$-invariant.
Next we show we may assume 
\begin{align}\label{cpteta}
\tilde \eta_{g,L}=\varsigma\tilde \eta_{g,R}\varsigma, 
\end{align}
for $\tilde\eta_{g,\sigma}$ 
 in (5.21) \cite{2df}.
Recall that
\begin{align}
\varphi_L\hat\otimes \varphi_R\sim_{q.e.} 
\omega_{p_{H_U\cap C_{\theta''}}}\circ\tau^{\ao g}\lmk\tilde\eta_{g,L}\hat\otimes \tilde\eta_{g,R}\rmk,
\end{align}
corresponding to (5.19), (5.20), (5.21) of \cite{2df}.
For a state $\varphi$ on $\caA_{H^U\cap{C_{\theta''}}}$, we denote by
$\varphi^{\varsigma}$ the state given by $\varphi^\varsigma(A):=\varphi\varsigma(A^*)$, $A\in\caA_{H^U\cap{C_{\theta''}}}$.
Note that if states $\varphi_1,\varphi_2$ on $\caA_{H^U\cap{C_{\theta''}}}$ are quasi-equivalent,
then $\varphi_1^\varsigma$ and $\varphi_2^\varsigma$ are also quasi-equivalent.
From this we have
\begin{align}
\begin{split}
&\omega_{p_{H_U\cap C_{\theta''}}}\circ\tau^{\ao g}\lmk\tilde\eta_{g,L}\hat\otimes \tilde\eta_{g,R}\rmk\\
&\sim_{q.e.}\varphi_L\hat\otimes \varphi_R
=\lmk \varphi_L\hat\otimes \varphi_R\rmk^{\varsigma}\\
&\sim_{q.e.}\lmk
\omega_{p_{H_U\cap C_{\theta''}}}\circ\tau^{\ao g}\lmk\tilde\eta_{g,L}\hat\otimes \tilde\eta_{g,R}\rmk
\rmk^\varsigma
=
\omega_{p_{H_U\cap C_{\theta''}}}\circ\tau^{-\ao g}\lmk\hat\eta_{g,L}\hat\otimes \hat\eta_{g,R}\rmk
\end{split}
\end{align}
In the last equality, we set
\begin{align}
\hat\eta_{g,L}:=\varsigma \tilde \eta_{g,R}\varsigma,\quad 
\hat\eta_{g,R}:=\varsigma \tilde \eta_{g,L}\varsigma,
\end{align}
and used $\varsigma\tau=\tau^{-1}\varsigma$ and the $\varsigma$-invariance of $\omega_{p_{H_U\cap C_{\theta''}}}$.
Because 
$\omega_{p_{H_U\cap C_{\theta''}}}\circ\tau^{-\ao g}=\omega_{p_{H_U\cap C_{\theta''}}}\circ\tau^{\ao g}$,
we then obtain
\begin{align}
\omega_{p_{H_U\cap C_{\theta''}}}\circ\tau^{\ao g}\lmk\tilde\eta_{g,L}\hat\otimes \tilde\eta_{g,R}\rmk
\simeq
\omega_{p_{H_U\cap C_{\theta''}}}\circ\tau^{\ao g}\lmk\hat\eta_{g,L}\hat\otimes \hat\eta_{g,R}\rmk.
\end{align}
From Lemma 4.1 \cite{2df}, this implies
\begin{align}
\omega_{p_{H_U\cap H_\sigma\cap C_{\theta''}}}
\lmk \tilde\eta_{g,\sigma}\hat\eta_{g,\sigma}^{-1}\rmk_{\ao g}
\simeq \omega_{p_{H_U\cap H_\sigma\cap C_{\theta''}}},\quad
\sigma=L,R.
\end{align}
(Recall Definition 3.3 \cite{2df} for the notation.)
Therefore, we conclude
\begin{align}
\omega_{p_{H_U\cap C_{\theta''}}}\circ\tau^{\ao g}\lmk\tilde\eta_{g,L}\hat\otimes \tilde\eta_{g,R}\rmk
\simeq
\omega_{p_{H_U\cap C_{\theta''}}}\circ\tau^{\ao g}\lmk\hat \eta_{g,L}\hat\otimes \tilde\eta_{g,R}\rmk.
\end{align}
In other words, we may choose $\tilde \eta_{g,L}$ so that
$\tilde\eta_{g,L}=\varsigma \tilde\eta_{g,R}\varsigma$, proving the claim (\ref{cpteta}).
Furthermore, in the proof of Proposition 5.2 \cite{2df}, from the construction of
the automorphic equivalence, $Z_\sigma$ can also be taken so that
$Z_L=\varsigma Z_R\varsigma$. 

Let
\begin{align}\label{bbs}
\alpha^{-1}\tau\alpha=\tau\lmk\zeta_L^1\hat\otimes \zeta_R^1\rmk\circ\inn
\end{align}
as in Lemma 3.2 \cite{2df}.
We also set $\zeta_\sigma^0:=\id$.
We then have (see (5.24), (5.25) \cite{2df})
\begin{align}\label{ringo}
\omega\beta_g^U\simeq 
\omega\alpha^{-1}\tau^{\ao g}\lmk\tilde \eta_{gL}Z_L\hat\otimes \tilde \eta_{gR}Z_R\rmk\alpha
=\omega\alpha^{-1}\tau^{\ao g}\alpha\alpha^{-1}\lmk\tilde \eta_{gL}Z_L\hat\otimes \tilde \eta_{gR}Z_R\rmk\alpha
\simeq 
\omega\tau^{\ao g} \lmk \zeta_L^{\ao g}\eta_{gL}'\hat\otimes \zeta_R^{\ao g}\eta_{gR}'\rmk,
\end{align}
with 
$\eta_{gL}'=\varsigma \eta_{gR}'\varsigma$.

Multiplying (\ref{bbs}) by $\varsigma$ from left and right we obtain
\begin{align}
\alpha^{-1}\tau^{-1}\alpha=\tau^{-1}\lmk\zeta_L^{-1}\hat\otimes \zeta_R^{-1}\rmk\circ\inn,
\end{align}
with
\begin{align}
\zeta_L^{-1}:=\varsigma \zeta_R^{1}\varsigma,
\quad \zeta_R^{-1}:=\varsigma \zeta_L^{1}\varsigma.
\end{align}
We then have
\begin{align}\label{ichigo}
\begin{split}
\omega\beta_g^U
=\lmk \omega\beta_g^U\rmk^{\varsigma}
\simeq 
\lmk \omega\tau^{\ao g} \lmk \zeta_L^{\ao g}\eta_{gL}'\hat\otimes \zeta_R^{\ao g}\eta_{gR}'\rmk\rmk^{\varsigma}\\
=
\omega\tau^{-\ao g} \lmk \zeta_L^{-\ao g}\eta_{gL}'\hat\otimes \zeta_R^{-\ao g}\eta_{gR}'\rmk.
\end{split}
\end{align}
Hence setting
\begin{align}
\eta_{g,L}^{1}:=\zeta_L^{\ao g}\eta_{gL}',\quad
\eta_{g,R}^{1}:=\zeta_R^{\ao g}\eta_{gR}'\quad
\eta_{g,L}^{-1}:=\zeta_L^{-\ao g}\eta_{gL}',\quad
\eta_{g,R}^{-1}:=\zeta_R^{-\ao g}\eta_{gR}',
\end{align}
(from (\ref{ringo} and (\ref{ichigo})) we have
\begin{align}
\omega\beta_g^U
\simeq 
\omega\tau^{\ao g\epsilon} \lmk \eta_{g,L}^{\epsilon}\hat\otimes \eta_{g,R}^{\epsilon }\rmk,\quad
\epsilon=\pm 1
\end{align}
and 
\begin{align}\label{sakura}
\varsigma\eta_{g,L}^{\epsilon}\varsigma=\eta_{g,R}^{-\epsilon},\quad
\epsilon=\pm 1
.
\end{align}
Hence we obtain $\lmk\eta_{g\sigma}^\epsilon\rmk\in I(\omega,\theta)$,
satisfying (\ref{sakura}).
Because of the $\varsigma$-invariance of $\omega^{(0)}$,
there is an antilinear unitary involution $J$ on $\caH_L\otimes\caH_R$
implementing $\varsigma$.
Because $\varsigma$ commutes with the grading $\Theta$, $J$
comuutes with $\Gamma_L\otimes\Gamma_R$
defined at the first paragraph of subsection 5.3 \cite{2df}.
Let $W_g$ be a unitary such that
\begin{align}
\Ad\lmk W_g\rmk \circ\lmk\pi_L\alpha_L\hat \otimes \pi_R\alpha_R\rmk
=\lmk\pi_L\alpha_L\hat \otimes \pi_R\alpha_R\rmk \circ\Upsilon
\beta_g^U\lmk \eta_{gL}^1 \hat\otimes \eta_{gR}^1 \rmk^{-1}
\tau^{-\ao{g}} \Upsilon^{-1},
\end{align}
which is given in Lemma 5.7 \cite{2df}.
Then we have
\begin{align}
\Ad\lmk J W_gJ\rmk \circ\lmk\pi_L\alpha_L\hat \otimes \pi_R\alpha_R\rmk
=\lmk\pi_L\alpha_L\hat \otimes \pi_R\alpha_R\rmk \circ\Upsilon
\beta_g^U\lmk \eta_{gL}^{-1} \hat\otimes \eta_{gR}^{-1} \rmk^{-1}
\tau^{\ao{g}} \Upsilon^{-1}.
\end{align}
Therefore, we may choose $W_g^\epsilon$ in $\IP\lmk \omega,\alpha,\theta, (\eta_{g\sigma}^\epsilon), (\alpha_L,\alpha_R,\Upsilon)\rmk$ (Definition 5.8 of \cite{2df}) so that
\begin{align}
W_g^{-\epsilon}=J W_g^\epsilon J,\quad \epsilon=\pm 1.
\end{align}
For this $W_g^\epsilon$, $b_g^\epsilon$ in Definition 5.8 \cite{2df} satisfies
\begin{align}\label{nashi}
b_g:=b_g^{-\epsilon}=b_g^\epsilon.
\end{align}
Let $ u_\sigma (g,h)$ be a unitary such that 
\begin{align}\label{u39}
\Ad\lmk u_\sigma (g,h)\rmk\pi_\sigma\alpha_\sigma
=\pi_\sigma\alpha_\sigma
\bgu{g}\sigma\lmk\eta_{g\sigma}^1\rmk^{-1}
\bgu{h}\sigma\lmk \lmk \eta_{h\sigma}^{(-1){^{\ao g}}}\rmk_{-\ao{g}}\rmk^{-1}
\eta_{gh,\sigma}^1 \lmk \bgu{gh}\sigma\rmk^{-1},
\end{align}
given by Lemma 5.6 \cite{2df}.
We have 
\begin{align}
\Ad\lmk W_g^{1} W_h^{(-1)^{a(g)}} \lmk W_{gh}^{1}\rmk ^*\rmk\circ\lmk\pi_L\alpha_L\hat \otimes \pi_R\alpha_R\rmk
=\Ad\lmk u_L (g,h)\otimes u_R (g,h)\Gamma_R^{\partial u_L (g,h)}\rmk
\circ\lmk\pi_L\alpha_L\hat \otimes \pi_R\alpha_R\rmk.
\end{align}
Here $\partial u_L (g,h)$ is the grading of $u_L (g,h)$.
Because $\lmk\pi_L\alpha_L\hat \otimes \pi_R\alpha_R\rmk
$ is irreducible, 
there is $\varsigma (g,h)\in\bbT$ such that
\begin{align}
\lmk W_g^1 W_h^{(-1)^{a(g)}} \lmk W_{gh}^1\rmk ^*\rmk
=\varsigma (g,h)
\lmk u_L (g,h)\otimes u_R (g,h)\Gamma_R^{\partial u_L (g,h)}\rmk.
\end{align}
Setting
\begin{align}
u_L^1 (g,h):=\varsigma (g,h)u_L (g,h),\quad
u_R^1 (g,h):=u_R (g,h),
\end{align}
we get
\begin{align}\label{wew}
\lmk W_g^1 W_h^{(-1)^{a(g)}} \lmk W_{gh}^1\rmk ^*\rmk
=
\lmk u_L^1 (g,h)\otimes u_R^1 (g,h)\Gamma_R^{\partial u_L^1 (g,h)}\rmk.
\end{align}
We have
\begin{align}
\Ad\lmk\unit\otimes u_R^1 (g,h)\rmk\lmk\pi_L\alpha_L\hat\otimes \pi_R\alpha_R\rmk
=\pi_L\alpha_L\hat\otimes \lmk\pi_R\alpha_R\circ \bgu{g}R\lmk\eta_{gR}^1\rmk^{-1}
\bgu{h}R\lmk \lmk \eta_{hR}^{(-1){^{\ao g}}}\rmk_{-\ao{g}}\rmk^{-1}
\eta_{gh,R}^1 \lmk \bgu{gh}R\rmk^{-1}\rmk
\end{align}
and
\begin{align}
\Ad\lmk u_L^1 (g,h)\otimes \Gamma_R^{\partial u_L(g,h)}\rmk\lmk\pi_L\alpha_L\hat\otimes \pi_R\alpha_R\rmk
=\lmk\pi_L\alpha_L\circ \bgu{g}L\lmk\eta_{gL}^1\rmk^{-1}
\bgu{h}L\lmk \lmk \eta_{hL}^{(-1){^{\ao g}}}\rmk_{-\ao{g}}\rmk^{-1}
\eta_{gh,L}^1 \lmk \bgu{gh}L\rmk^{-1}\rmk
\hat\otimes \pi_R\alpha_R.
\end{align}
(Recall Lemma C.1 \cite{2df}.) 
We then have
\begin{align}
\Ad\lmk J \lmk \unit\otimes u_R^1 (g,h)\rmk J\rmk\lmk\pi_L\alpha_L\hat\otimes \pi_R\alpha_R\rmk
=\lmk\pi_L\alpha_L\circ \bgu{g}L\lmk\eta_{gL}^{-1}\rmk^{-1}
\bgu{h}L\lmk \lmk \eta_{hL}^{-(-1){^{\ao g}}}\rmk_{\ao{g}}\rmk^{-1}
\eta_{gh,L}^{-1} \lmk \bgu{gh}L\rmk^{-1}\rmk
\hat\otimes \pi_R\alpha_R.
\end{align}
and
\begin{align}
\Ad\lmk J\lmk u_L^1 (g,h)\otimes \Gamma_R^{\partial u_L(g,h)}\rmk J\rmk \lmk\pi_L\alpha_L\hat\otimes \pi_R\alpha_R\rmk
=\pi_L\alpha_L\hat\otimes \lmk\pi_R\alpha_R\circ \bgu{g}R\lmk\eta_{gR}^{-1}\rmk^{-1}
\bgu{h}R\lmk \lmk \eta_{hR}^{-(-1){^{\ao g}}}\rmk_{\ao{g}}\rmk^{-1}
\eta_{gh,R}^{-1} \lmk \bgu{gh}R\rmk^{-1}\rmk.
\end{align}
Therefore, setting
\begin{align}
 u_L^{-1} (g,h)\otimes \Gamma_R^{\partial u_L^{-1}(g,h)}
:=J \lmk \unit\otimes u_R^1 (g,h)\rmk J,\\
\unit\otimes u_R^{-1} (g,h):=
J\lmk u_L^1 (g,h)\otimes \Gamma_R^{\partial u_L(g,h)}\rmk J
\end{align}
we obtain $\{u_\sigma^\epsilon(g,h)\}$ in $\IP\lmk \omega,\alpha,\theta, (\eta_{g\sigma}^\epsilon), (\alpha_L,\alpha_R,\Upsilon)\rmk$.
Note that for this $u_\sigma^\epsilon(g,h)$, $\kappa_\sigma^\epsilon(g,h)$ in
Definition 5.8 \cite{2df} satisfies
\begin{align}\label{bibi}
\kss L{-\epsilon}(g,h) =\kss R{\epsilon}(g,h).
\end{align}
Applying $\Ad J$ to (\ref{wew}),
\begin{align}
\lmk W_g^{-1} W_h^{-(-1)^{a(g)}} \lmk W_{gh}^{-1}\rmk ^*\rmk
=
\lmk u_L^{-1} (g,h)\otimes u_R^{-1} (g,h)\Gamma_R^{\partial u_L^{-1} (g,h)}\rmk\cdot (-1)^{\kss L{-1}(g,h)\kss R{-1}(g,h)}
.
\end{align}
Hence we obtain
\begin{align}
\lmk W_g^{\epsilon} W_h^{(-1)^{a(g)}\epsilon } \lmk W_{gh}^\epsilon\rmk ^*\rmk
=
\lmk u_L^{\epsilon} (g,h)\otimes u_R^{\epsilon} (g,h)\Gamma_R^{\partial u_L^\epsilon (g,h)}\rmk
\cdot (-1)^{\kss L{-1}(g,h)\kss R{-1}(g,h)\frac{1-\epsilon}2}.
\end{align}
From this we obtain
\begin{align}\label{bbbkk}
b_g+b_h+b_{gh}=\kss L\epsilon(g,h)+\kss R\epsilon(g,h).
\end{align}

Recall from Lemma 5.10 \cite{2df} we have 
\begin{align}\label{lem41eq}
\begin{split}
&W_g^\epsilon\lmk\unih L\otimes \uss R {\eg g \epsilon}(h,k)\rmk{W_g^\epsilon}^*
\lmk \unih L\otimes \uss R \epsilon (g,hk)\rmk
=\css R\epsilon (g,h,k) \lmk\unih L \otimes \uss R\epsilon (g,h) \uss R\epsilon (gh,k)\rmk\\
&W_g^\epsilon \lmk \uss L {\eg{g}\epsilon}(h,k)\otimes \Gamma_R^{\kss L {\eg g \epsilon}(h,k)}\rmk{W_g^\epsilon}^*
\lmk \uss  L\epsilon (g,hk)\otimes \Gamma_R^{\kss L \epsilon (g,hk)}\rmk
=\css L \epsilon (g,h,k) \uss L \epsilon (g, h)\uss L \epsilon (gh, k)\otimes \Gamma_R^{\kss L \epsilon (g,h)+\kss L \epsilon (gh,k)}.
\end{split}
\end{align}
Taking $\Ad J$, we obtain
\begin{align}\label{lem41e}
\begin{split}
&W_g^{-\epsilon} \lmk \uss L {-\eg{g}{\epsilon}}(h,k)\otimes \Gamma_R^{\kss L {-\eg g {\epsilon}}(h,k)}\rmk{W_g^{-\epsilon}}^*
\lmk \uss  L{-\epsilon} (g,hk)\otimes \Gamma_R^{\kss L {-\epsilon} (g,hk)}\rmk\\
&=\overline{\css R \epsilon (g,h,k)} \uss L {-\epsilon} (g, h)\uss L {-\epsilon} (gh, k)\otimes \Gamma_R^{\kss L {-\epsilon} (g,h)+\kss L {-\epsilon} (gh,k)}.
\\
&W_g^{-\epsilon}\lmk\unih L\otimes \uss R {-\eg g \epsilon}(h,k)\rmk{W_g^{-\epsilon}}^*
\lmk \unih L\otimes \uss R {-\epsilon} (g,hk)\rmk
=\overline{\css L{\epsilon} (g,h,k)} \lmk\unih L \otimes \uss R{-\epsilon} (g,h) \uss R{-\epsilon} (gh,k)\rmk\\
&={\css R{-\epsilon} (g,h,k)} \lmk\unih L \otimes \uss R{-\epsilon} (g,h) \uss R{-\epsilon} (gh,k)\rmk.
\end{split}
\end{align}
Hence for our $W_g^\epsilon$, $u_\sigma^\epsilon(g,h)$, 
we have
\begin{align}\label{budou}
\overline{\css L{\epsilon} (g,h,k)}={\css R{-\epsilon} (g,h,k)}.
\end{align}

Note from (\ref{lem41eq}) and (2.8) \cite{2df}, we have
\begin{align}
W_g^\epsilon \lmk \uss L {\eg{g}\epsilon}(h,k)\otimes \Gamma_R^{\kss L {\eg g \epsilon}(h,k)}\rmk{W_g^\epsilon}^*
=\tilde v_L(g,h,k)\otimes  \Gamma_R^{\kss L {\eg g \epsilon}(h,k)}
\end{align}
with some unitary $\tilde v_L(g,h,k)$ on $\caH_L$.
Using this, we have
\begin{align}\label{ume}
\begin{split}
&W_g^\epsilon\lmk\unih L\otimes \uss R {\eg g \epsilon}(h,k)\rmk{W_g^\epsilon}^*
\lmk \unih L\otimes \uss R \epsilon (g,hk)\rmk
W_g^\epsilon \lmk \uss L {\eg{g}\epsilon}(h,k)\otimes \Gamma_R^{\kss L {\eg g \epsilon}(h,k)}\rmk{W_g^\epsilon}^*
\lmk \uss  L\epsilon (g,hk)\otimes \Gamma_R^{\kss L \epsilon (g,hk)}\rmk\\
&=W_g^\epsilon\lmk\unih L\otimes \uss R {\eg g \epsilon}(h,k)\rmk{W_g^\epsilon}^*
\lmk \unih L\otimes \uss R \epsilon (g,hk)\rmk
\lmk \tilde v_L(g,h,k)\otimes  \Gamma_R^{\kss L {\eg g \epsilon}(h,k)}\rmk
\lmk \uss  L\epsilon (g,hk)\otimes \Gamma_R^{\kss L \epsilon (g,hk)}\rmk\\
&=(-1)^{\kss L {\eg g \epsilon}(h,k)\cdot \kss R \epsilon (g,hk)}\\
&W_g^\epsilon\lmk\unih L\otimes \uss R {\eg g \epsilon}(h,k)\rmk{W_g^\epsilon}^*
\lmk \tilde v_L(g,h,k)\otimes  \Gamma_R^{\kss L {\eg g \epsilon}(h,k)}\rmk
\lmk \unih L\otimes \uss R \epsilon (g,hk)\rmk
\lmk \uss  L\epsilon (g,hk)\otimes \Gamma_R^{\kss L \epsilon (g,hk)}\rmk\\
&=(-1)^{\kss L {\eg g \epsilon}(h,k)\cdot \kss R \epsilon (g,hk)}\\
&W_g^\epsilon\lmk\unih L\otimes \uss R {\eg g \epsilon}(h,k)\rmk{W_g^\epsilon}^*
W_g^\epsilon \lmk \uss L {\eg{g}\epsilon}(h,k)\otimes \Gamma_R^{\kss L {\eg g \epsilon}(h,k)}\rmk{W_g^\epsilon}^*
\lmk \unih L\otimes \uss R \epsilon (g,hk)\rmk
\lmk \uss  L\epsilon (g,hk)\otimes \Gamma_R^{\kss L \epsilon (g,hk)}\rmk\\
&=(-1)^{\kss L {\eg g \epsilon}(h,k)\cdot \kss R \epsilon (g,hk)}\\
&W_g^\epsilon\lmk  \uss L {\eg{g}\epsilon}(h,k)\otimes \uss R {\eg g \epsilon}(h,k)\Gamma_R^{\kss L {\eg g \epsilon}(h,k)}\rmk
{W_g^\epsilon}^*
\lmk \uss  L\epsilon (g,hk)\otimes \uss R \epsilon (g,hk)\Gamma_R^{\kss L \epsilon (g,hk)}\rmk\\
&=(-1)^{\kss L {\eg g \epsilon}(h,k)\cdot \kss R \epsilon (g,hk)}\\
&
W_g^\epsilon\lmk   W_h^{{\eg{g}\epsilon}}W_k^{(-1)^{\ao h}{\eg{g}\epsilon}}
\lmk W_{hk}^{{\eg{g}\epsilon}}\rmk^*\rmk
{W_g^\epsilon}^*
W_g^\epsilon W_{hk}^{(-1)^{\ao g}\epsilon }\lmk W_{ghk}^\epsilon\rmk^*\\
&(-1)^{\kss L{-1}(h,k)\kss R{-1}(h,k)\frac {1-(-1)^{a(g)}\epsilon}2}
(-1)^{\kss L{-1}(g,hk)\kss R{-1}(g,hk)\frac {1-\epsilon}2}
\\
&=(-1)^{\kss L {\eg g \epsilon}(h,k)\cdot \kss R \epsilon (g,hk)}(-1)^{\kss L{-1}(h,k)\kss R{-1}(h,k)\frac {1-(-1)^{a(g)}\epsilon}2}
(-1)^{\kss L{-1}(g,hk)\kss R{-1}(g,hk)\frac {1-\epsilon}2}
\\
&
W_g^\epsilon  W_h^{{\eg{g}\epsilon}}W_k^{(-1)^{\ao h}{\eg{g}\epsilon}}
 \lmk W_{ghk}^\epsilon\rmk^*.
\end{split}
\end{align}
On the other hand, we have
\begin{align}\label{mokuren}
\begin{split}
&\css R\epsilon (g,h,k) \lmk\unih L \otimes \uss R\epsilon (g,h) \uss R\epsilon (gh,k)\rmk
\css L \epsilon (g,h,k)\lmk  \uss L \epsilon (g, h)\uss L \epsilon (gh, k)\otimes \Gamma_R^{\kss L \epsilon (g,h)+\kss L \epsilon (gh,k)}\rmk\\
&=(-1)^{\kss L \epsilon (g,h)\cdot \kss R\epsilon (gh,k)}
\lmk \css R\epsilon (g,h,k) \css L \epsilon (g,h,k)\rmk
\lmk \uss L \epsilon (g, h)\uss L \epsilon (gh, k)\otimes \uss R\epsilon (g,h) \Gamma_R^{\kss L \epsilon (g,h)}
\uss R\epsilon (gh,k)\Gamma_R^{\kss L \epsilon (gh,k)}\rmk\\
&=(-1)^{\kss L \epsilon (g,h)\cdot \kss R\epsilon (gh,k)}
\lmk \css R\epsilon (g,h,k) \css L \epsilon (g,h,k)\rmk
W_g^\epsilon W_h^{\eg g \epsilon}\lmk W_{gh}^\epsilon \rmk^*
W_{gh}^\epsilon W_k^{\eg {gh} \epsilon}\lmk W_{ghk}^\epsilon \rmk^*\\
&(-1)^{\kss L{-1}(g,h)\kss R{-1}(g,h)\frac{1-\epsilon}2}
(-1)^{\kss L{-1}(gh,k)\kss R{-1}(gh,k)\frac{1-\epsilon}2}\\
&=(-1)^{\kss L \epsilon (g,h)\cdot \kss R\epsilon (gh,k)}(-1)^{\kss L{-1}(g,h)\kss R{-1}(g,h)\frac{1-\epsilon}2}
(-1)^{\kss L{-1}(gh,k)\kss R{-1}(gh,k)\frac{1-\epsilon}2}
\lmk \css R\epsilon (g,h,k) \css L \epsilon (g,h,k)\rmk\\
&W_g^\epsilon W_h^{\eg g \epsilon}W_k^{\eg {gh} \epsilon}\lmk W_{ghk}^\epsilon \rmk^*.
\end{split}
\end{align}
Note from (\ref{lem41eq}) that (\ref{ume}) and (\ref{mokuren}) are equal.

Hence we get 
\begin{align}
\begin{split}
&(-1)^{\kss L{-1}(g,h)\kss R{-1}(g,h)\frac{1-\epsilon}2}
(-1)^{\kss L{-1}(gh,k)\kss R{-1}(gh,k)\frac{1-\epsilon}2}
(-1)^{\kss R {-\epsilon} (g,h)\cdot \kss R\epsilon (gh,k)}
\lmk \css R\epsilon (g,h,k) \overline{\css R {-\epsilon} (g,h,k)}\rmk\\
&=(-1)^{\kss L{-1}(g,h)\kss R{-1}(g,h)\frac{1-\epsilon}2}
(-1)^{\kss L{-1}(gh,k)\kss R{-1}(gh,k)\frac{1-\epsilon}2}
(-1)^{\kss L \epsilon (g,h)\cdot \kss R\epsilon (gh,k)}
\lmk \css R\epsilon (g,h,k) \css L \epsilon (g,h,k)\rmk\\
&=(-1)^{\kss L {\eg g \epsilon}(h,k)\cdot \kss R \epsilon (g,hk)}(-1)^{\kss L{-1}(h,k)\kss R{-1}(h,k)\frac {1-(-1)^{a(g)}\epsilon}2}
(-1)^{\kss L{-1}(g,hk)\kss R{-1}(g,hk)\frac {1-\epsilon}2}\\
&=(-1)^{\kss R {-\eg g \epsilon}(h,k)\cdot \kss R \epsilon (g,hk)}(-1)^{\kss L{-1}(h,k)\kss R{-1}(h,k)\frac {1-(-1)^{a(g)}\epsilon}2}
(-1)^{\kss L{-1}(g,hk)\kss R{-1}(g,hk)\frac {1-\epsilon}2}.
\end{split}
\end{align}
Here we used (\ref{bibi}), (\ref{budou}) for the first equation, and 
(\ref{ume})$=$(\ref{mokuren}) for the second equation.
Setting $\epsilon=1$ in this equation, we have 
\begin{align} \label{cici}
\begin{split}
&\lmk \css R 1 (g,h,k) \overline{\css R {-1} (g,h,k)}\rmk\\
&=
(-1)^{\kss L{-1}(h,k)\kss R{-1}(h,k)\frac {1-(-1)^{a(g)}}2}
(-1)^{\kss R {-\eg g 1}(h,k)\cdot \kss R 1 (g,hk)}
(-1)^{\kss R {-1} (g,h)\cdot \kss R1 (gh,k)}\\
&=
(-1)^{\kss L{-1}(h,k)\kss R{-1}(h,k)\frac {1-(-1)^{a(g)}}2}
(-1)^{\kss R {\eg g 1}(h,k)\cdot \kss R 1 (g,hk)}
(-1)^{\kss R {1} (g,h)\cdot \kss R1 (gh,k)}\\
&(-1)^{\lmk \kss R {-\eg g  1}(h,k)-  \kss R {\eg g 1}(h,k)\rmk\cdot \kss R 1 (g,hk)}
(-1)^{\lmk \kss R {-1} (g,h)- \kss R {1} (g,h)\rmk \cdot \kss R1 (gh,k)}\\
&=
(-1)^{\kss L{-1}(h,k)\kss R{-1}(h,k)\frac {1-(-1)^{a(g)}}2}
e^{i\frac \pi 2 \lmk
\lmk \kss R {\eg g 1}(h,k)+ \kss R 1 (g,hk)\rmk^2
-\lmk \kss R {\eg g 1}(h,k)\rmk^2- \lmk \kss R 1 (g,hk)\rmk^2
\rmk}\\&e^{-i\frac \pi 2\lmk
\lmk
\kss R {1} (g,h)+ \kss R1 (gh,k)
\rmk^2
-\lmk
\kss R {1} (g,h)\rmk^2
-\lmk  \kss R1 (gh,k)
\rmk^2
\rmk
}\\
&(-1)^{\lmk \kss R {-\eg g 1}(h,k)-  \kss R {\eg g 1}(h,k)\rmk\cdot \kss R 1 (g,hk)}
(-1)^{\lmk \kss R {-1} (g,h)- \kss R {1} (g,h)\rmk \cdot \kss R1(gh,k)}\\
&=(-1)^{\kss L{-1}(h,k)\kss R{-1}(h,k)\frac {1-(-1)^{a(g)}}2}
e^{i\frac \pi 2 \lmk
-\lmk \kss R {\eg g 1}(h,k)\rmk- \lmk \kss R 1 (g,hk)\rmk
\rmk}\\
&e^{i\frac \pi 2\lmk
\lmk
\kss R {1} (g,h)\rmk
+\lmk  \kss R1 (gh,k)
\rmk
\rmk
}
(-1)^{\lmk d^1b (h,k)\rmk\cdot \kss R 1 (g,hk)}
(-1)^{\lmk d^1b (g,h)\rmk \cdot \kss R1 (gh,k)}.
\end{split}
\end{align}

Here, from the third equality, with a bit abuse of notation, we set $\kappa_\sigma^\epsilon=0\in\bbZ$/
$\kappa_\sigma^\epsilon=1\in\bbZ$
if $\kappa_\sigma^\epsilon=0\in\bbZ_2$/$\kappa_\sigma^\epsilon=1\in\bbZ_2$.
Note that the difference between 
$\lmk \kss R {\eg g 1}(h,k)+ \kss R 1 (g,hk)\rmk^2$
and 
$\lmk
\kss R {1} (g,h)+ \kss R1 (gh,k)
\rmk^2
$
is $0$, $4$ or $-4$.
We used this fact and (\ref{bibi}) at the fourth equality.
Note that $\lmk d_a^1b (g,h)\rmk^{1}=\lmk d_a^1b (g,h)\rmk^{-1}$
because of (\ref{nashi}). We denote this value by $d^1b (g,h)$.

Now we would like to show that 
$(c_R,\kappa_R,\kappa_L,b,a)\sim(\hat c, \kappa,\kappa,0,a)$
for some $[\hat c,\kappa,a]\in \Hom(\Omega_3^{spin}(BG), \bbR\setminus \bbZ)$.
Setting
\begin{align}
m^\epsilon(g):=
\left\{
\begin{gathered}
0,\quad \epsilon=1\\
b_g,\quad \epsilon=-1
\end{gathered}
\right.,
\end{align}
and
\begin{align}
\begin{split}
\kappa(g,h):=\kappa_R(g,h)+d^1_a m(g,h)
\end{split}
\end{align}
we have
\begin{align}
\begin{split}
&\kappa^{+1}(g,h)=\lmk d_a^1m(g,h)\rmk^1+\kss R 1(g,h)
=b_ha(g)+\kss R 1(g,h)\\
&\kappa^{-1}(g,h)=\lmk d_a^1m(g,h)\rmk^{-1}+\kss R {-1}(g,h)
=\lmk d_a^1m(g,h)\rmk^{-1}+\kss L{1}(g,h)\\
&=\lmk d_a^1m(g,h)\rmk^{-1}+d^1b(g,h)+\kss R{1}(g,h)
\quad =b_ha(g)+\kss R 1(g,h),
\end{split}
\end{align}
using (\ref{bibi}) and (\ref{bbbkk}).
Hence we have
\begin{align}\label{ksim}
\kappa^{+1}(g,h)=\kappa^{-1}(g,h).
\end{align}
With this notation, from (\ref{cici}), we have
\begin{align}
\begin{split}&\lmk \css R 1 (g,h,k) \overline{\css R {-1} (g,h,k)}\rmk\\
&=
(-1)^{\kss L{-1}(h,k)\kss R{-1}(h,k)\frac {1-(-1)^{a(g)}}2}
e^{i\frac \pi 2 \lmk
-\lmk \kss R {\eg g 1}(h,k)\rmk- \lmk \kss R 1 (g,hk)\rmk
\rmk}
e^{i\frac \pi 2\lmk
\lmk
\kss R {1} (g,h)\rmk+\lmk  \kss R1 (gh,k)
\rmk
\rmk
}
\\
&(-1)^{\lmk d^1b (h,k)\rmk\cdot \lmk d^1_am  (g,hk)\rmk^1}
(-1)^{\lmk d^1b (g,h)\rmk \cdot \lmk d^1_am (gh,k)\rmk^{1}}
(-1)^{\lmk d^1b (h,k)\rmk\cdot \kappa (g,hk)}
(-1)^{\lmk d^1b (g,h)\rmk \cdot \kappa (gh,k)}.
\end{split}
\end{align}

Setting
\begin{align}
\begin{split}
\kappa'(g,h):=\kappa_L(g,h)-d^1_a m(g,h)
=\kappa_R(g,h)+d^1_a b(g,h)+d^1_am(g,h)=\kappa(g,h)+d^1_a b(g,h)
\end{split}
\end{align}
 we have
\begin{align}
(\kappa')^{+1}(g,h)=(\kappa')^{-1}(g,h).
\end{align}
Furthermore, set
\begin{align}
c(g,h,k):=(-1)^{\kappa_L (g,h)\cdot m^{\ao {gh}}(k)}(-1)^{m(g)\cdot \kappa^{\ao g}(h,k) } 
c_R(g,h,k). 
\end{align}
We then have 
\begin{align}
(c,\kappa,\kappa',b,a)\sim_{\PD} (c_R, \kappa_R,\kappa_L,b,a).
\end{align}
Now we set
\begin{align}
\begin{split}
&\tilde c(g,h,k):=(-1)^{b_g\cdot \kappa(h,k)} c(g,h,k).
\end{split}
\end{align}
(Recall $\kappa^{\ao g}(h,k)=\kappa(h,k)$ by (\ref{ksim}))
From Lemma 2.4 \cite{2df}, we have 
\begin{align}
\begin{split}
(c,\kappa,\kappa',b,a)\sim (\tilde c, \kappa,\kappa,0,a).
\end{split}
\end{align}
Now we would like to rewrite
\begin{align}
\begin{split}
&\tilde c(g,h,k)
=(-1)^{b_g\cdot \kappa(h,k)} 
(-1)^{\kappa_L (g,h)\cdot m^{\ao {gh}}(k)}(-1)^{m(g)\cdot \kappa^{\ao g}(h,k) } 
c_R(g,h,k).
\end{split}
\end{align}
Because
\begin{align}
\begin{split}
\kss L{}(g,h)
=d^1b(g,h)+d^1_a m(g,h)+\kappa(g,h),
\end{split}
\end{align}
we have
\begin{align}
\begin{split}
&\tilde c(g,h,k)
=(-1)^{b_g\cdot \kappa(h,k)} 
(-1)^{\lmk d^1b(g,h)+d^1_a m(g,h)+\kappa(g,h)\rmk \cdot m^{\ao {gh}}(k)}(-1)^{m(g)\cdot \kappa(h,k) } 
c_R(g,h,k).
\end{split}
\end{align}
Note also that
\begin{align}
\begin{split}
&d_a^1m(g,h)^\epsilon=\left\{
\begin{gathered}
a(g) b_h,\quad \quad \quad \epsilon=1,\\
d_a^1b(g,h)-a(g) b_h\quad \quad \quad \epsilon=-1
\end{gathered}
\right.\\
&=d_a^1b(g,h)\frac{1-\epsilon}{2}+ a(g)b_h.
\end{split}
\end{align}
Set 
\begin{align}\label{suika}
\begin{split}
\sigma(g,h)^\epsilon:=\left\{
\begin{gathered}
1,\quad\quad\quad\quad \epsilon=1,\\
 e^{i\frac \pi 2\kss R 1(g,h)} (-1)^{\kappa(g,h)\lmk b_g+b_h\rmk},\quad\quad\quad\quad \epsilon=-1
\end{gathered}
\right.
\end{split}
\end{align}
and
\begin{align}\label{kuri}
\begin{split}
\eta^\epsilon(g,h):=\left\{
\begin{gathered}
1,\quad\quad\quad \epsilon=1,\\
e^{-i\frac \pi4 \lmk b_g+b_h-b_{gh}\rmk},\quad\quad\quad \epsilon=-1\end{gathered}
\right..
\end{split}
\end{align}
Here, with a bit abuse of notation, we set $\kappa_\sigma^\epsilon=0\in\bbZ$/
$\kappa_\sigma^\epsilon=1\in\bbZ$ in (\ref{suika})
if $\kappa_\sigma^\epsilon=0\in\bbZ_2$/$\kappa_\sigma^\epsilon=1\in\bbZ_2$.
Similarly, in (\ref{kuri}) we set
$b=0\in\bbZ$/
$b=1\in\bbZ$
if $b=0\in\bbZ_2$/$b=1\in\bbZ_2$.
We claim 
\begin{align}
\begin{split}
\tilde c^1(g,h,k)\overline{\tilde c^{-1}(g,h,k)}
=\frac{\lmk d^2_a\sigma\eta(g,h,k)\rmk^1}{\lmk d^2_a\sigma\eta(g,h,k)\rmk^{-1}}.
\end{split}
\end{align}
In fact, we have
\begin{align}
\begin{split}
&\tilde c^1(g,h,k)\cdot \overline{\tilde c^{-1}(g,h,k)}\\
&=
(-1)^{\lmk d^1b(g,h)\rmk \cdot b_k}
(-1)^{b_h a(g)a(gh) b_k+ \lmk d^1b(g,h) - b_h a(g)\rmk b_k (1-a(gh))}(-1)^{\lmk \kappa(g,h)\rmk \cdot b_k}
(-1)^{b(g)\cdot \kappa(h,k) } 
c_R^1(g,h,k)\overline{c_R^{-1}(g,h,k)}\\
&=
(-1)^{\lmk d^1b(g,h)\rmk \cdot b_k}
(-1)^{b_h a(g)b_k+ \lmk d^1b(g,h) \rmk b_k (1-a(gh))}(-1)^{\lmk \kappa(g,h)\rmk \cdot b_k}
(-1)^{b(g)\cdot \kappa(h,k) } 
c_R^1(g,h,k)\overline{c_R^{-1}(g,h,k)}
\\
&=
(-1)^{b_h a(g)b_k+ \lmk d^1b(g,h) \rmk b_k a(gh)}(-1)^{\lmk \kappa(g,h)\rmk \cdot b_k}
(-1)^{b(g)\cdot \kappa(h,k) } 
c_R^1(g,h,k)\overline{c_R^{-1}(g,h,k)}
\\
&=
(-1)^{b_h a(g)b_k+ \lmk d^1b(g,h) \rmk b_k a(gh)}
(-1)^{\kss L{-1}(h,k)\kss R{-1}(h,k)\frac {1-(-1)^{a(g)}}2}
e^{i\frac \pi 2 \lmk
-\lmk \kss R {\eg g 1}(h,k)\rmk- \lmk \kss R 1 (g,hk)\rmk
\rmk}e^{i\frac \pi 2\lmk
\lmk
\kss R {1} (g,h)\rmk+\lmk  \kss R1 (gh,k)
\rmk
\rmk
}
\\
&(-1)^{\lmk d^1b (h,k)\rmk\cdot \lmk d^1_am  (g,hk)^1\rmk}
(-1)^{\lmk d^1b (g,h)\rmk \cdot \lmk d^1_am (gh,k)^1\rmk}
(-1)^{\lmk \kappa(g,h)\rmk \cdot b_k}
(-1)^{b(g)\cdot \kappa(h,k) }\\
&(-1)^{\lmk b_h-b_k+b_{hk}\rmk\cdot \kappa (g,hk)}
(-1)^{\lmk b_g-b_h+b_{gh}\rmk \cdot \kappa (gh,k)}
\\
&=
(-1)^{b_h a(g)b_k+ \lmk d^1b(g,h) \rmk b_k a(gh)}
(-1)^{\kss L{-1}(h,k)\kss R{-1}(h,k)\frac {1-(-1)^{a(g)}}2}
e^{i\frac \pi 2 \lmk
-\lmk \kss R {\eg g 1}(h,k)\rmk- \lmk \kss R 1 (g,hk)\rmk
\rmk}e^{i\frac \pi 2\lmk
\lmk
\kss R {1} (g,h)\rmk+\lmk  \kss R1 (gh,k)
\rmk
\rmk
}
\\
&(-1)^{\lmk d^1b (h,k)\rmk\cdot \lmk d^1_am  (g,hk)^1\rmk}
(-1)^{\lmk d^1b (g,h)\rmk \cdot \lmk d^1_am (gh,k)^1\rmk}
(-1)^{\lmk \kappa(g,h)-\kappa (g,hk) \rmk \cdot b_k}
(-1)^{b(g)\cdot \lmk \kappa (gh,k)+\kappa(h,k) \rmk}\\
&
(-1)^{b_h\lmk \kappa (g,hk)- \kappa (gh,k)\rmk}
(-1)^{\lmk b_{hk}\rmk\cdot \lmk \kappa (g,hk)\rmk}
(-1)^{\lmk b_{gh}\rmk \cdot \kappa (gh,k)}
\\
&=
(-1)^{b_h a(g)b_k+ \lmk d^1b(g,h) \rmk b_k a(gh)}
(-1)^{\kss L{-1}(h,k)\kss R{-1}(h,k)\frac {1-(-1)^{a(g)}}2}
\\
&e^{i\frac \pi 2 \lmk
-\lmk \kss R {\eg g 1}(h,k)\rmk- \lmk \kss R 1 (g,hk)\rmk
\rmk}e^{i\frac \pi 2\lmk
\lmk
\kss R {1} (g,h)\rmk+\lmk  \kss R1 (gh,k)
\rmk
\rmk
}
\\
&(-1)^{\lmk d^1b (h,k)\rmk\cdot \lmk d^1_am  (g,hk)^1\rmk}
(-1)^{\lmk d^1b (g,h)\rmk \cdot \lmk d^1_am (gh,k)^1\rmk}\\
&(-1)^{\lmk \kappa(h,k)+ \kappa(gh,k)  \rmk \cdot b_k}
(-1)^{b(g)\cdot \lmk \kappa(g,h)+\kappa(g,hk)  \rmk}
(-1)^{b_h\lmk \kappa(g,h)+\kappa(h,k)\rmk}
(-1)^{\lmk b_{hk}\rmk\cdot \lmk \kappa (g,hk)\rmk}
(-1)^{\lmk b_{gh}\rmk \cdot \kappa (gh,k)}
\\
&=\frac{d^2_a\sigma(g,h,k)^1}{d^2_a\sigma(g,h,k)^{-1}}
(-1)^{b_h a(g)b_k+ \lmk d^1b(g,h) \rmk b_k a(gh)}
(-1)^{\kss L{-1}(h,k)\kss R{-1}(h,k)\frac {1-(-1)^{a(g)}}2}
\\
&e^{i\frac \pi 2 \lmk
-\lmk \kss R {\eg g 1}(h,k)\rmk
\rmk}
\overline{e^{i\frac \pi 2 \lmk
\lmk \kss R {1}(h,k)\rmk
\rmk}}^{a(g)}
\\
&(-1)^{\lmk d^1b (h,k)\rmk\cdot \lmk d^1_am  (g,hk)\rmk^1}
(-1)^{\lmk d^1b (g,h)\rmk \cdot \lmk d^1_am (gh,k)\rmk^1}\\
&=\frac{d^2_a\sigma(g,h,k)^1}{d^2_a\sigma(g,h,k)^{-1}}
(-1)^{b_h a(g)b_k+ \lmk d^1b(g,h) \rmk b_k a(gh)}
(-1)^{\kss R{1}(h,k)\kss R{-1}(h,k)\frac {1-(-1)^{a(g)}}2}
\\
&e^{i\frac \pi 2 \lmk
-\lmk \kss R {\eg g 1}(h,k)\rmk
\rmk}
\overline{e^{i\frac \pi 2 \lmk
\lmk \kss R {1}(h,k)\rmk
\rmk}}^{a(g)}
(-1)^{\lmk d^1b (h,k)\rmk\cdot \lmk d^1_am  (g,hk)\rmk^1}
(-1)^{\lmk d^1b (g,h)\rmk \cdot \lmk d^1_am (gh,k)\rmk^1}.
\end{split}
\end{align}
Here in the fourth equality we substituted (\ref{cici}).
We used $2$-cocycle relation for the sixth equality
In the seventh equality we used a notation
${\bar c}^a=c,\bar c$ for $a=0,1$.
In the last equality we used (\ref{bibi}).
Note that
\begin{align}
\begin{split}
(-1)^{\kss R{1}(h,k)\kss R{-1}(h,k)}
=e^{i\frac\pi 2\lmk -\lmk\kss L{-1}(h,k)+\kss R{-1}(h,k)\rmk^2+\kss L{-1}(h,k)+\kss R{-1}(h,k)\rmk }
=e^{-i\frac\pi 2\lmk b_h-b_k+b_{hk}\rmk^2 }
e^{i\frac\pi 2\lmk \kss R{1}(h,k)+\kss R{-1}(h,k)\rmk }\\
\end{split}
\end{align}
Here, with a bit abuse of notation, we set $\kss\sigma \epsilon=0,1\in\bbZ$
if $\kss\sigma \epsilon=0,1\in\bbZ_2$.
Analogously,  we set $b_g=0,1\in\bbZ$
if $b_g=0,1\in\bbZ_2$.
Note that the difference between $\lmk b_h-b_k+b_{hk}\rmk^2$ and
$\lmk\kss L{-1}(h,k)+\kss R{-1}(h,k)\rmk^2$ are $0,-4,4$
that the last equality holds.
From this we get
%
\begin{align}
\begin{split}
&(-1)^{\kss R{1}(h,k)\kss R{-1}(h,k)\frac {1-(-1)^{a(g)}}2}
e^{i\frac \pi 2 \lmk
-\lmk \kss R {\eg g 1}(h,k)\rmk
\rmk}
\overline{e^{i\frac \pi 2 \lmk
\lmk \kss R {1}(h,k)\rmk
\rmk}}^{a(g)}\\
&=e^{-i\frac\pi 2\lmk b_h-b_k+b_{hk}\rmk^2 a(g)}
\end{split}
\end{align}
Substituting this we get
\begin{align}
\begin{split}&\tilde c^1(g,h,k)\cdot \overline{\tilde c^{-1}(g,h,k)}\\
&=\frac{d^2_a\sigma(g,h,k)^1}{d^2_a\sigma(g,h,k)^{-1}}
(-1)^{b_h a(g)b_k+ \lmk d^1b(g,h) \rmk b_k a(gh)}\\
&
e^{-i\frac\pi 2\lmk b_h-b_k+b_{hk}\rmk^2 a(g)}
\\
&(-1)^{\lmk d^1b (h,k)\rmk\cdot \lmk d^1_am  (g,hk)\rmk^1}
(-1)^{\lmk d^1b (g,h)\rmk \cdot \lmk d^1_am (gh,k)\rmk^1}\\
&=\frac{d^2_a\sigma(g,h,k)^1}{d^2_a\sigma(g,h,k)^{-1}}
(-1)^{b_h a(g)b_k+ \lmk b_g+b_h+b_{gh} \rmk b_k a(gh)}\\
&
e^{-i\frac\pi 2\lmk b_h+b_k+b_{hk}-2b_hb_k+2b_hb_{hk}-2b_k b_{hk}\rmk a(g)}
\\
&(-1)^{\lmk b_h+b_k+b_{hk}\rmk\cdot \lmk a(g) b_{hk} \rmk}
(-1)^{\lmk b_g+b_h+b_{gh}\rmk \cdot \lmk a(gh)b_k \rmk}\\
&=\frac{d^2_a\sigma(g,h,k)^1}{d^2_a\sigma(g,h,k)^{-1}}
(-1)^{b_h a(g)b_k+ \lmk b_g+b_h+b_{gh} \rmk b_k a(gh)}\\
&
(-1)^{\lmk b_hb_k+b_hb_{hk}+b_k b_{hk}\rmk a(g)}
e^{-i\frac\pi 2\lmk b_h+b_k+b_{hk}\rmk a(g)}
\\
&(-1)^{\lmk b_h+b_k+b_{hk}\rmk\cdot \lmk a(g) b_{hk} \rmk}
(-1)^{\lmk b_g+b_h+b_{gh}\rmk \cdot \lmk a(gh)b_k \rmk}\\
&=\frac{d^2_a\sigma(g,h,k)^1}{d^2_a\sigma(g,h,k)^{-1}}
(-1)^{\lmk b_hb_{hk}+b_k b_{hk}\rmk a(g)}
e^{-i\frac\pi 2\lmk b_h+b_k+b_{hk}\rmk a(g)}
\\
&(-1)^{\lmk b_h+b_k+b_{hk}\rmk\cdot \lmk a(g) b_{hk} \rmk}\\
&=\frac{d^2_a\sigma(g,h,k)^1}{d^2_a\sigma(g,h,k)^{-1}}
e^{-i\frac\pi 2\lmk b_h+b_k-b_{hk}\rmk a(g)} \\
&=\frac{d^2_a\sigma\eta(g,h,k)^1}{d^2_a\sigma\eta(g,h,k)^{-1}}
\end{split}
\end{align}
Set
\begin{align}
\begin{split}
\hat c(g,h,k):=d_a^2\lmk \overline{\sigma\eta}\rmk(g,h,k)\tilde c(g,h,k).
\end{split}
\end{align}
Then we have
\begin{align}
\begin{split}
 (\tilde c, \kappa,\kappa,0,a)\sim_{\PD}  (\hat c, \kappa,\kappa,0,a).
\end{split}
\end{align}
We can see that $[\hat c,\kappa,a]\in \Hom(\Omega_3^{spin}(BG), \bbR\setminus \bbZ)$
from (\ref{ksim}) and 
\begin{align}
\begin{split}
&\frac{\hat c(g,h,k)^1}{\hat c(g,h,k)^{-1}}\\
&=\frac{d_a^2\lmk \overline{\sigma\eta}\rmk(g,h,k)^1}{d_a^2\lmk \overline{\sigma\eta}\rmk(g,h,k)^{-1}}
\tilde c^1(g,h,k)\cdot \overline{\tilde c^{-1}(g,h,k)}\\
&=\frac{d_a^2\lmk \overline{\sigma\eta}\rmk(g,h,k)^1}{d_a^2\lmk \overline{\sigma\eta}\rmk(g,h,k)^{-1}}
\frac{d^2_a\sigma\eta(g,h,k)^1}{d^2_a\sigma\eta(g,h,k)^{-1}}\\
&=1.
\end{split}
\end{align}

\noindent{\bf Acknowledgment.}
{
The author is grateful to Yuji Tachikawa for a stimulating discussion and helpful comments. 
This work was supported by JSPS KAKENHI Grant Number 19K03534 and 22H01127.
It was also supported by JST CREST Grant Number JPMJCR19T2.
}\\\\
\noindent{DATA AVAILABILITY
The data that support the findings of this study are available within the article.}

\appendix

\end{document}